\documentclass{iopjournal}

\usepackage{amsmath,amssymb}    
\usepackage{graphicx}           
\usepackage{hyperref}           
\usepackage{siunitx}            
\usepackage{booktabs}           

\usepackage{lscape}
\usepackage{xcolor}
\usepackage[numbers,sort&compress]{natbib}
\begin{document}

\articletype{Paper} 

\title{Harnessing data-driven methods for precise model independent event shape estimation in relativistic heavy-ion collisions}

\author{Dipankar Basak$^{1,2}$\orcid{0000-0002-7878-6256}, H. Hushnud$^3$\orcid{0000-0002-8203-4671} and Kalyan Dey$^{1,*}$\orcid{0000-0002-4633-2946}}

\affil{$^1$Department of Physics, Bodoland University, Kokrajhar, 783370, Assam, India}

\affil{$^2$Department of Physics, Kokrajhar University, Kokrajhar, 783370, Assam, India}

\affil{$^3$Department of Physics, Koneru Lakshmaiah Education Foundation, Vaddeswaram, Guntur, 522502, Andhra Pradesh, India}

\affil{$^*$Corresponding author}

\email{kalyan.dey@buniv.edu.in}


\begin{abstract}
This study demonstrates the application of supervised machine learning (ML) techniques to distinguish between isotropic and jet-like event topologies in heavy-ion collisions via the spherocity observable. State-of-the-art ML algorithms, optimized through systematic hyperparameter tuning, are employed to predict both traditional transverse spherocity $S_{0}$ and unweighted transverse spherocity $S_{0}^{p_{\rm T}=1}$ directly from raw event data. Moreover, the results from this study demonstrated that our approach remains largely model-independent, underscoring its potential applicability in future experimental heavy-ion physics analyses.
\end{abstract}

\section{Introduction}
\label{introduction}

Event-shape engineering has recently emerged as a powerful method to explore collective-like phenomena and probe possible medium effects in high-multiplicity proton-proton (pp) collisions at ultra-relativistic energies \cite{Khuntia_2021,ESE5,ESE6,ESE7,sahoo}. Event classification according to final-state geometrical features (jet-like versus isotropic) enables a more differential analysis and offers advantages over conventional methods. Transverse spherocity $(S_{0})$, \cite{ORTIZ201578, Oliva2022,Ortiz2024, Tripathy2021} is one such  event‑shape observable that effectively distinguishes between hard, jet‑dominated (“jetty”) and soft, isotropic events in hadronic and heavy-ion collisions. Transverse spherocity has been successfully used to study various properties of high-multiplicity pp collisions at LHC energies \cite{ESE1,ESE2,ESE3,Khatun_2020}. This technique has recently been extended to heavy‑ion collisions, providing refined control over initial-state geometries at fixed centrality and opening new avenues for studying medium effects in a multi‑differential context \cite{Mallick_2021,Prasad2022,Mallick2022,PRASAD2025100110}. This event‑shape selection allows detailed, multi‑differential studies of how global observables such as the mean transverse momentum $\langle p_{\rm T} \rangle$, particle spectra, and flow‑like correlations depend simultaneously on charged‑particle multiplicity and event geometry.

In recent years, data-driven approaches using machine learning have been successfully used across many areas of high-energy physics, including event classification, particle identification, jet flavor tagging, etc. \cite{ML_1,ML_2,ML_3,ML_4,ML_5}. These methods have consistently demonstrated substantial improvements over traditional analysis techniques. The motivation for employing a machine-learning estimator of transverse spherocity is twofold. Experimentally, the direct, event-by-event computation of $S_{0}$ requires full charged-track reconstruction over full azimuth and is sensitive to local inefficiencies, dead regions, and acceptance gaps; these effects can introduce topology-dependent biases that are difficult to correct on an event-by-event basis. Computationally, the standard $S_{0}$ calculation involves a minimization over azimuthal directions, which can be costly for large-scale analyses or online selection. By contrast, global observables such as $dN_{\rm ch}/d\eta$, $\langle m_{T}\rangle$, and $v_{2}$ (or the proxy $\langle\cos 2\phi\rangle$) are routinely measured with established correction procedures. An ML mapping from these robust inputs to $S_{0}$ thus provides a practical, indirect estimator of event topology that is resilient to missing or noisy track-level information and is suitable for fast classification tasks.\\ 

Mallick et al. \cite{Mallick_sph_ML} previously estimated transverse spherocity ($S_0$) using gradient-boosted decision trees, relying on total charged-particle multiplicity, mean transverse momentum, and transverse-region multiplicity, which requires region-based selection and thus additional preprocessing. Building on this framework, our work focuses on Au+Au collisions at RHIC energies and introduces a physically motivated set of features: the mean transverse mass, $\langle m_{T} \rangle$, and a proxy for elliptic flow, $\langle \cos(2\phi) \rangle$, which are directly sensitive to event topology \cite{Prasad2022, Mallick2022} and can be extracted from reconstructed tracks with minimal preprocessing. This approach avoids the need for region-based selections while retaining sensitivity to the jetty-isotropic transition. In addition to revisiting the choice of input observables, we systematically benchmark a broad class of machine-learning architectures including Polynomial regression, $K$‑nearest neighbors, Decision Trees regressor, Extra‑Trees regressor, Light Gradient Boosting Machine, and shallow neural networks networks to predict both the conventional transverse spherocity $S_0$ and unweighted transverse spherocity $S_0^{p_{\rm T}=1}$ for Au+Au collisions at $\sqrt{s_{\rm NN}}=200$ GeV. This comparative study enables us to evaluate the dependence of the results on the choice of ML architecture, identify the most reliable predictors, and establish the robustness of the proposed observable set for spherocity estimation.  These additions provide a more comprehensive methodological assessment and complement the framework introduced in Ref.~\cite{Mallick_sph_ML}.
Unlike Ref.~\cite{Mallick_sph_ML}, which considered only the traditional definition of spherocity, the present work estimates both the traditional and unweighted forms of $S_{0}$, thereby offering a more complete characterization of event topology. The unweighted spherocity, by treating all particles on equal footing irrespective of their transverse momentum, enhances sensitivity to the global geometric structure of the event and mitigates bias from high-$p_{\rm T}$ particles. This distinction is particularly relevant for disentangling isotropic and jet-like features in heavy-ion collisions.

ML models trained on simulations from a specific event generator often suffer from model dependence, limiting their predictive power across generators and constraining their applicability to experimental data. This motivates the development of model-independent, data-driven methods for reliable event-shape estimation in heavy-ion collisions. To address model dependence, the present study will perform comprehensive hyperparameter tuning across diverse ML families, aiming to ensure robustness and generalizability across multiple event generators, thereby making it well-suited for future experimental analyses.

The paper is structured as follows: Section \ref{TS} provides a brief introduction to transverse spherocity followed by an overview of the various heavy-ion event generators in Section \ref{Generator}. Section \ref{ML} presents the machine learning regression models employed in this study. In Section \ref{Method}, we describe the data generation and preprocessing procedures, model optimization, and the performance metrics used. A detailed discussion of the results is presented in Section \ref{Results}, and the paper concludes with a summary in Section \ref{Summary}.

\section{Transverse Spherocity}
\label{TS}

Transverse spherocity is an event shape observable that is used to distinguish the events based on their azimuthal event topology. It is defined relative to a unit vector $(\hat{n})$ that minimizes the ratio \cite{Banfi2010}:

\begin{equation}
S_0 = \frac{\pi^2}{4}{\left(\frac{\sum_i\mid \vec{p}_{{\rm T}_{i}} \times \hat{n} \mid}{\sum_i\vec{p}_{{\rm T}_{i}}}\right)}^2
\label{eq:1}.
\end{equation}

\begin{figure}[h]
\centering 
\includegraphics[width=0.5\textwidth]{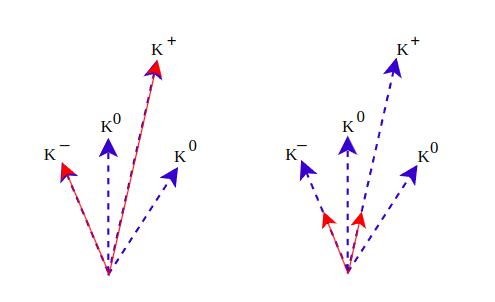}
\caption{Illustration of weights utilization (represented by red arrows) during computation of $S_0$ (left) and $S_0^{p_{\rm T}=1}$ (right) in an jetty event.}\label{fig1}
\end{figure}

The summations are performed on all primary charged particles. Transverse spherocity is a normalized quantity whose values ranges from $0$ to $1$. The $S_0 \rightarrow 0$ corresponds to events (jetty events) mainly dominated by a single hard scattering with back-to-back jet structures  while  $S_0 \rightarrow 1$  corresponds to events (isotropic events) where particles are isotropically distributed due to soft QCD interactions \cite{Tripathy2021}.
Due to the $p_{\rm T}$-weight in the definition of $S_0$ estimator and as it consider only charged particles, there is always a possibility of charged-to-neutral bias in the computation of $S_0$ \cite{Acharya2024}.  That means in the calculation of transverse spherocity, a single high-$p_{\rm T}$ charged particle will have a large weight compared to several low-$p_{\rm T}$ neutral particles that may isotropically distributed in the same event. To reduce this possible charged-vs-neutral bias, a new estimator known as Unweighted Transverse Spherocity ($S_0^{p_{\rm T}=1}$)  was introduced \cite{Nassirpour:2848793,Acharya2024} where the transverse momentum ($p_{\rm T}$) of each charged particles were normalized to 1 ($p_{\rm T} = 1$ GeV/c) in Eq.~\ref{eq:1}. With same weight assigned to each particle, the  calculation of  transverse spherocity becomes more robust against individual tracks with large $p_{\rm T}$. Fig.~\ref{fig1} illustrates how $p_{\rm T}$ weights are considered in the calculation of  $S_0$ and $S_0^{p_{\rm T}=1}$. The unweighted transverse spherocity is expressed as:

\begin{equation}
S_0^{p_{\rm T}=1} = \frac{\pi^2}{4}{\left(\frac{\sum_i\mid \hat{p}_{{\rm T}_{i}} \times \hat{n} \mid}{N_{trks}}\right)}^2
\label{eq:2}.
\end{equation}

where $N_{trks}$ is the total number of charged particles. From now on, $S_0$ and $S_0^{p_{\rm T}=1}$ will be referred as conventional spherocity and unweighted spherocity respectively. Events with at least five charged hadrons ($p_{\rm T}>0.2\ \mathrm{GeV}/c$, $|\eta|<1$) were used to compute both $S_0$ and $S_0^{p_{\rm T}=1}$.

\section{\label{Generator}Event Generation}

In this study, we have employed four different event generators such as \texttt{AMPT}, \texttt{UrQMD}, \texttt{PYTHIA-Angantyr} and \texttt{EPOS4} to simulate heavy-ion collision events. Brief description of these models are given below:

\subsection{\label{AMPT}AMPT}
The \texttt{AMPT} (A Multi-Phase Transport) \cite{AMPT} model is a widely used Monte Carlo event generator used to simulate heavy-ion collisions at relativistic energies. It has four key components which models different phases (i) initial conditions based on the heavy-ion jet interaction generator model (\texttt{HIJING}) \cite{HIJING}, (ii) parton interaction by Zhang’s parton cascade (ZPC) model \cite{ZPC}, (iii) hadronization is either via Lund string fragmentation model \cite{Lund} (in default version) or via quark coalescence model \cite{He:2017} (in string model version) and (iv) finally, the hadronic re-scattering phase using ART (A Relativistic Transport) model \cite{ART}. In this study we have used string melting version of \texttt{AMPT}. The data generated with \texttt{AMPT} were used to train and evaluate the performance of the ML models in predicting the values of $S_0$ and $S_0^{p_{\rm T}=1}$. We also used few other heavy-ion event generators which were utilized to check the model independency of ML models trained with \texttt{AMPT} data by evaluating their performance.

\subsection{\label{UrQMD}UrQMD} 
Ultra-relativistic quantum molecular dynamics (\texttt{UrQMD}) is a microscopic model typically used for simulating nucleus-nucleus collisions at ultra-relativistic beam energies \cite{BASS1998255, M_Bleicher_1999}. It is a transport model that propagates hadrons covariantly along classical trajectories, providing a numerical solution via Monte Carlo methods to the relativistic Boltzmann equation. Its key mechanisms encompass a stochastic treatment of two-body scatterings, the dynamics of color string production and fragmentation, and the formation and decay of resonances. It is one of successful transport models for describing heavy-ion collisions at wide energy ranges ($\sqrt{s_{\rm NN}}=2$ GeV to 200 GeV) \cite{Urqmd_energy_1,Urqmd_energy_2,Urqmd_energy_3,Urqmd_energy_4}. The \texttt{UrQMD} model incorporates a comprehensive set of hadronic states, including both ground states and their resonances. The collision term includes 55 baryonic and 32 mesonic species, along with their corresponding antiparticles and all isospin-projected states. In this study, we employed the \texttt{UrQMD} version 3.3 with the default parameters.

\subsection{\label{Pythia}PYTHIA8 Angantyr}
\texttt{PYTHIA8} \cite{PYTHIA8} is a Monte Carlo event generator that was initially designed to study small collision systems such as $e^{+}e^{-}$, $pp$ and $p\bar{p}$ collisions. Recently efforts have been made to use \texttt{PYTHIA8} to describe ultra-relativistic heavy-ion collisions. That is done by incorporating the Angantyr model \cite{Angantyr} in \texttt{PYTHIA8} which uses the direct extrapolation of high-energy pp collisions without considering the formation of a hot thermalised medium during simulation. The model is inspired by the Fritiof model \cite{Fritiof} and the concept of wounded nucleons and was successful in reproducing  of general final-state properties of both $p\rm A$ and $\rm AA$ collisions \cite{pythia_1,pythia_2,pythia_3}.

\subsection{\label{Epos}EPOS4}
\texttt{EPOS4} \cite{EPOS4, EPOS4_2} is a multipurpose event generator based on Monte Carlo framework that simulates high-energy $pp$, $p\rm A$ and $\rm AA$ collisions, including both initial- and final-state dynamics. It is the latest version of \texttt{EPOS} \cite{EPOS} (\textbf{E}nergy conserving quantum mechanical approach, based on \textbf{P}artons, parton ladders, strings, \textbf{O}ff-shell remnants, and \textbf{S}plitting of parton ladders). The main new development in \texttt{EPOS4} over its previous versions is to  accommodate  factorization and saturation in a parallel multiple scattering  scenario~\cite{EPOS4_N}. The \texttt{EPOS} model has several phases of evolution: The initial stage based on  Gribov-Regge multiple scattering theory \cite{Gribov–Regge}, the other main components of EPOS4 are the core-corona, hydrodynamical evolution, hadronization and the hadronic cascade. \texttt{EPOS4} has been successfully applied to study various collisions systems at different energies like Au+Au collisions at RHIC energies, $pp$ collisions at $\sqrt{s}=13.6$ TeV, $p\rm Ne$ collisions at $\sqrt{s}=68.5$ GeV, O+O collisions at $\sqrt{s_{\rm NN}}=7$ TeV \cite{EPOS_4, EPOS_1,EPOS_3, EPOS_2}. In this work we have used the 4.0.3 version of \texttt{EPOS4}.

\section{\label{ML}Machine Learning Algorithms}
Data-driven methods such as deep learning (DL) and machine learning (ML) have the capability to  analyze and interpret complex datasets by extracting patterns in data to make predictions or decisions without explicit programming. In this study, we have used several ML regression algorithms to estimate the value of transverse spherocity. A brief description of the methods is provided below.

\subsection{\label{PR}Polynomial Regression}
Polynomial Regression (PR) \cite{PR} is an extension of linear regression model that models the relationship between the independent variable $x$ and the dependent variable $y$ as a polynomial of $n^\textrm{th}$ degree: 
\begin{equation}
y = \beta_0 + \beta_1x +\beta_2x^2 + ....... + \beta_nx^n
\label{eq:3}.
\end{equation}

 where $\beta_0$, $\beta_1$, ...., $\beta_n$ are the coefficients of the polynomial terms. The goal of the algorithm is to determine the coefficients so that the difference between the predicted values and the actual values in the training data is minimized. In this algorithm, first the original input features are converted into polynomial features and then training is done on the transformed features using any linear model. In this work we have used Ridge regression model \cite{ridge}, a regularized version of linear regression model.

\subsection{\label{DTR}Decision Tree Regressor}
A Decision Tree Regressor (DTR) is a machine learning algorithm that uses a tree-like structure to predict continuous target variables \cite{DTR}. It recursively splits the data into subsets based on the decision rules derived from the input features to reduce the variance in the target variable within each subset. This process continues, creating new nodes and branches until a stopping criterion is met. The model predicts a continuous value at the leaf node which is corresponds to the average of the target values in that node.

\subsection{\label{ETR}Extra-Trees Regressor} 
The Extra-Trees or Extremely Randomized Trees (ETR) \cite{ETR} is a machine learning algorithm operates similarly to Random Forest Regressor. Random Forest \cite{RFR} is based on ensemble learning where several randomly generated decision trees are used to make a prediction. It integrates the independent predictions of all decision trees to produce final prediction of the model. Due to the use of multiple decision trees, the prediction performance and generalization ability of the model are enhanced. In ETR algorithm, instead of searching for best features like Random Forest, it randomly selects features to prevent overfitting and improves its generalization performance.

\subsection{\label{KNN}K-Nearest Neighbor}
The K-Nearest Neighbors (KNN) algorithm is a supervised, non-parametric machine learning technique used for both classification and regression problems \cite{KNN}. It is a simple algorithm that predicts the value of a target variable based on a fixed number (K) of its closest (nearest) neighbors in the feature space.  It uses distance metric (e.g., Euclidean distance) to calculate its nearest neighbors from the training data. The predicted value is the average of the numerical target of these neighbors.

\subsection{\label{LightGBM }Light Gradient Boosting Machine} 
Light Gradient Boosting Machine (LGBM) \cite{LightGBM} is an upgraded version of the  standard Gradient-boosting algorithm \cite{GBR}.  It is an ensemble-based machine learning approach can be used for both classification and regression problems. It creates a sequence of weak models, typically decision trees, to make predictions. Each new decision tree is trained to minimize the loss function  of the previous model using gradient descent optimization.  This approach iteratively reduces the model's prediction error.  The combined predictions of these decision trees are used to build a strong predictive model.

\subsection{\label{MLP}Multi-layer Perceptron Network} 
The multi-layer Perceptron Network (MLP) \cite{MLP} is a supervised feed-forward neural network used for both regression and classification. It has three types of layers: an input layer, single or multiple hidden layers, and an output layer. Each node of one layer is connected to each node of the next layer. All of nodes in the hidden and output layers consists of neurons with a non-linear activation function in addition to the input nodes. The output of the first layer becomes the input for the next layer and this process continues until the output layer is reached. The back-propagation learning technique \cite{backpro} is applied during the training of the MLP network. \\

The ML algorithms used in this study were successfully applied to both various classification and regression tasks in high-energy physics like PR in \cite{PR_Ex}, LGBM in \cite{Li_2020,PhysRevC.109.024604,PhysRevC.106.014901}, MLP in \cite{Xiang_2022,Basak2023,Galaktionov2023,BASAK2025123043}, KNN in \cite{PhysRevC.106.014901}, ETR in \cite{PhysRevC.106.014901}, DT in \cite{PhysRevC.106.014901} etc. All the ML-models were implemented using the \texttt{Scikit-Learn} toolboxes \cite{Scikit-learn} in python.

\begin{figure}
\centering
\includegraphics[width=\textwidth]{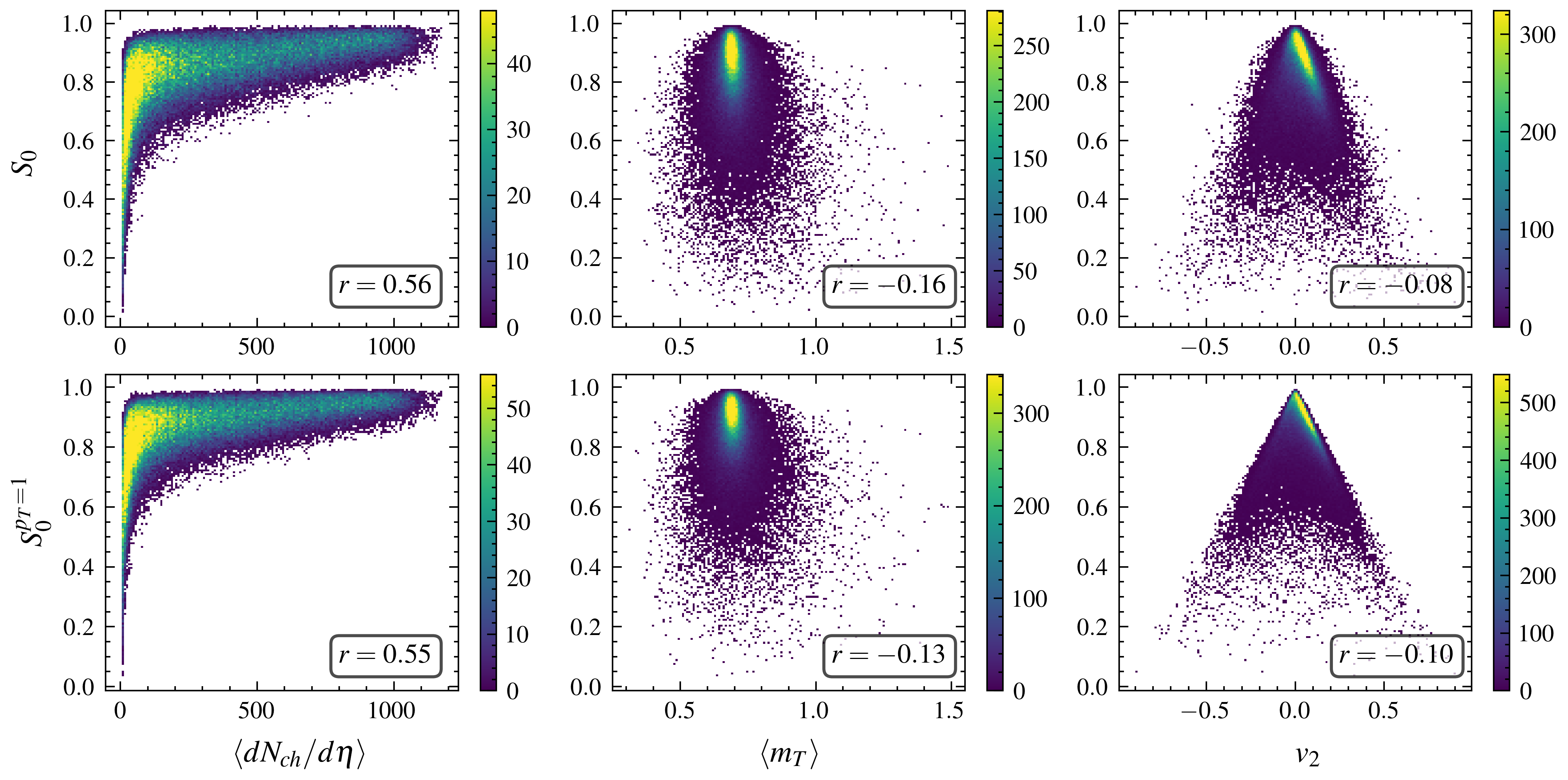}
\caption{Correlation plots between the input features and the target variables. The values of Pearson correlation coefficient $r$ are also shown.}\label{fig2}
\end{figure}

\section{\label{Method}Methodology}
\subsection{Data Preparation}

Microscopic Monte Carlo transport models provide a comprehensive framework for simulating the full evolution of heavy-ion collisions, encompassing all stages from the initial nuclear overlap through partonic and hadronic phases to kinetic freeze-out. In the present study, the string melting version of A Multi-Phase Transport (\texttt{AMPT}) model is employed to produce datasets for training our machine learning models. A total of 1.1 million minimum-bias Au+Au collision events at  $\sqrt{s_{\rm NN}}=200$ GeV were generated out of which 100 K events were kept aside as test data for inference.\\

The current study addresses an inverse problem aimed at determining the event shape through conventional and unweighted spherocity in heavy-ion collisions on an event-by-event basis. This challenge is approached through the application of machine learning-based regression techniques. Three experimentally measurable observables, namely charge particle multiplicity at mid-rapidity $\langle dN_{\rm ch}/d\eta \rangle$, the average transverse mass of charged particles $\langle m_{\rm T} \rangle$, and the elliptic flow parameter $\langle v_2 \rangle$ were used as inputs or features for training the ML models. Their selection is motivated by known physical connections to event topology. For instance, the mid-rapidity charged particle density $\langle dN_{\rm ch}/d\eta \rangle$, serves as a proxy for the entropy production, is known to correlate strongly with spherocity where with higher charged particle multiplicity classes, the spherocity distribution shifts towards higher values \cite{refId0}. Similarly, $\langle m_T \rangle$ values are known to be enhanced for jetty events \cite{Prasad2022}. Furthermore, the elliptic flow coefficient exhibits a pronounced anti-correlation with transverse spherocity, as isotropic events approach $\langle v_{2} \rangle \approx 0$ while jetty configurations yield the largest elliptic anisotropy \cite{Mallick2022}. Elliptic flow describes the anisotropy in the momentum distribution of the produced particles and is believed to sensitive to the initial state of heavy-ion collisions \cite{Elliptic}. Typically, $v_2$ is estimated by measuring the second harmonic coefficient in the Fourier expansion of the azimuthal distribution of particle transverse momentum and expressed as $v_2 = \langle \cos{[2(\phi-\psi)]}\rangle$, where $\phi$ is the azimuthal angle and $\psi$ is the reaction plane angle. It should be noted that the reaction plane angle $\psi$ cannot be measured in experiment directly. It is generally estimated using methods like event plane method \cite{event_plane}, azimuthal correlations method \cite{RPAzC}, via the feature of linear polarization of the coherent photoproduction process \cite{RPAzC2} etc. In the present investigation, the effect of reaction plane angle on the accuracy of the results will be discussed.

Fig.~\ref{fig2} presents scatter plots of each input observable against the target spherocity variable. Superimposed on the panels are the Pearson correlation coefficients, which indicate that the mid‐rapidity charged‐particle density exhibits the strongest positive correlation with both conventional and unweighted spherocity, $r\bigl[\langle \mathrm{d}N_{\mathrm{ch}}/\mathrm{d}\eta \rangle, S_0 (S_0^{p_T=1})\bigr] = 0.56 (0.55)$.
In contrast, the mean transverse mass and the elliptic flow coefficient show weaker negative correlations, $
r\bigl[\langle m_{\rm T} \rangle, S_0 (S_0^{p_T=1})\bigr] = -0.16(-0.13),
r\big[(v_{2}, S_0 (S_0^{p_T=1})\bigr] = -0.08  (-0.10)$ with $v_{2}$ displaying the smallest absolute correlation.

Feature scaling constitutes an essential preprocessing step in machine learning, as algorithms typically attain better convergence and performance when input variables share a common range \cite{geron2022hands}. In this work, we employ \texttt{StandardScaler} module from scikit-learn \cite{Scikit-learn} to standardize each feature to zero mean and unit variance, thereby ensuring all inputs lie on the same scale.

\subsection{\label{Tuning}Hyperparameter Optimization}

The performance of a ML algorithm significantly depends on its hyperparameters values. Thus proper set of hyperparameters are necessary to obtain an optimal model. Tuning hyperparameters can increase prediction accuracy for a particular dataset \cite{Hyper-parameter}. \texttt{GridSearchCV} (Grid Search and Cross Validation) is one of the technique in machine learning used for hyperparameter tuning \cite{GridsearchCV}. \texttt{GridSearchCV} checks the model for every possible combinations of parameters within a specified range using cross-validation and determines the optimal hyperparameter combination with the highest performance score for the model. In this paper, we used \texttt{GridSearchCV} with 5-fold cross validation to determine the best combination of hyperparameters for each ML algorithm and those combination were implemented during the training process. Table \ref{table1} shows hyperparameters which were optimized and the best parameters for each model. Other hyperparameters which were not optimized were kept at model default values.

\begin{landscape}
\begin{table}
\caption{Results of hyperparameter optimization.}
\centering
\begin{tabular}{l c c c c}
\hline
\textrm{ML-Model}& \textrm{Hyperparameters}& \textrm{Values and Ranges }&
\multicolumn{2}{c}{\textrm{Optimal Hyperparameters}}\\
    &   &   & $S_0$ & $S_0^{p_{\rm T}=1}$\\
\hline
PR & poly\underline{\hspace{0.2cm}}degree & [2, 3, 4, 5, 6, 7, 8] & 7 & 7 \\
    & ridge\underline{\hspace{0.2cm}}alpha & [0.01, 0.1, 1, 10, 100] & 10 & 10 \\
    & ridge\underline{\hspace{0.2cm}}solver & [`auto', `svd', `cholesky', `lsqr'] & `svd' & `svd'\\

DTR & criterion & [`squared\underline{\hspace{0.2cm}}error', `friedman\underline{\hspace{0.2cm}}mse', `absolute\underline{\hspace{0.2cm}}error', `poisson'] & `poisson' & `poisson' \\
    & max\underline{\hspace{0.2cm}}depth & [None, 5, 10, 15, 20] & 10 & 10 \\
    & max\underline{\hspace{0.2cm}}features & [None, `sqrt', `log2'] & None & None \\
    & min\underline{\hspace{0.2cm}}samples\underline{\hspace{0.2cm}}leaf & [1, 2, 4] & 4 & 4 \\  
    & min\underline{\hspace{0.2cm}}samples\underline{\hspace{0.2cm}}split & [2, 5, 10] & 10 & 10 \\
    
ETR & max\underline{\hspace{0.2cm}}depth & [None, 10, 20, 30] & 10 & 10 \\
    & min\underline{\hspace{0.2cm}}samples\_leaf & [1, 2, 4]  & 2 & 1 \\
    & min\underline{\hspace{0.2cm}}samples\_split & [2, 5, 10] & 10 & 5 \\
    & n\underline{\hspace{0.2cm}}estimators & [50, 100, 200] & 100 & 100 \\   

KNN & algorithm & [`auto', `ball\underline{\hspace{0.2cm}}tree', `kd\underline{\hspace{0.2cm}}tree', `brute'] & `auto' & `auto' \\
    & metric & [`euclidean', `manhattan', `minkowski'] & `manhattan' & `manhattan' \\
    & n\underline{\hspace{0.2cm}}neighbors & [1 - 100]  & 50 & 45\\
    & weights & [`uniform', `distance'] & `distance' & `distance' \\
 
LGBM & learning\underline{\hspace{0.2cm}}rate & [0.01, 0.1, 0.15, 0.2] & 0.01 & 0.05 \\
    & max\underline{\hspace{0.2cm}}depth & [3, 5, 7, 10] & 5 & 5\\
    & n\underline{\hspace{0.2cm}}estimators & [100, 200, 500,600] & 500 & 100 \\
    & num\underline{\hspace{0.2cm}}leaves &  [31, 50, 70] & 31 & 31 \\

MLP  & activation & [`relu', `tanh'] & `relu' & `relu' \\
    & alpha & [0.0001, 0.001, 0.01] & 0.001 & 0.001 \\
    & hidden\underline{\hspace{0.2cm}}layer\underline{\hspace{0.2cm}}sizes & [(50,), (100,), (50, 50), (100, 50),(150,100,50), (100,50,30)] & (100,) & (50,)\\
    & learning\underline{\hspace{0.2cm}}rate & [`constant', `adaptive'] & `constant' & `constant' \\
    & solver & [`adam', `lbfgs'] & `lbfgs' & `lbfgs' \\
\hline
\end{tabular}
\label{table1}
\end{table}
\end{landscape}

\subsection{\label{Performance}Performance Matrices}

The performance of all the models used here were evaluated by comparing the model predicted values with the actual values of spherocity. This was done by using three evaluation matrices namely mean absolute error (MAE), root mean squared error (RMSE) and coefficient of determination ($R^2$). MAE measures the mean of the absolute difference between the predicted and actual values whereas RMSE determines the square root of the average squared difference between the estimated values and the actual values. $R^2$ is a statistical metric which describes the goodness of fit of a regression model. These matrices are expressed as, 
\begin{equation}
\text{MAE} = \frac{1}{n}\sum_{i=1}^{n} \mid y_i^{t} - y_i^{p} \mid
\label{eq:4}.
\end{equation}

\begin{equation}
\text{RMSE} = \sqrt{\frac{1}{n}\sum_{i=1}^{n}{\left( y_i^{t} - y_i^{p} \right)}^2}
\label{eq:5}.
\end{equation}

\begin{equation}
R^2 = 1 - \frac{\sum_{i=1}^{n}{\left( y_i^{t} - y_i^{p} \right)}^2}{\sum_{i=1}^{n}{\left(y_i^t- \langle y^t \rangle \right)}^2}
\label{eq:6}.
\end{equation}

where $y_i^t$, $y_i^p$ and  $\langle y^t \rangle$ represent the true values, predicted values and the average of the true values of the target variable respectively. Smaller the value of MAE and RMSE, the higher the accuracy of the model whereas for $R^2$ it is opposite. An ideal model would achieve zero values of MAE and RMSE and $R^2 = 1$.

\begin{figure}
\centering
\includegraphics[width=\textwidth]{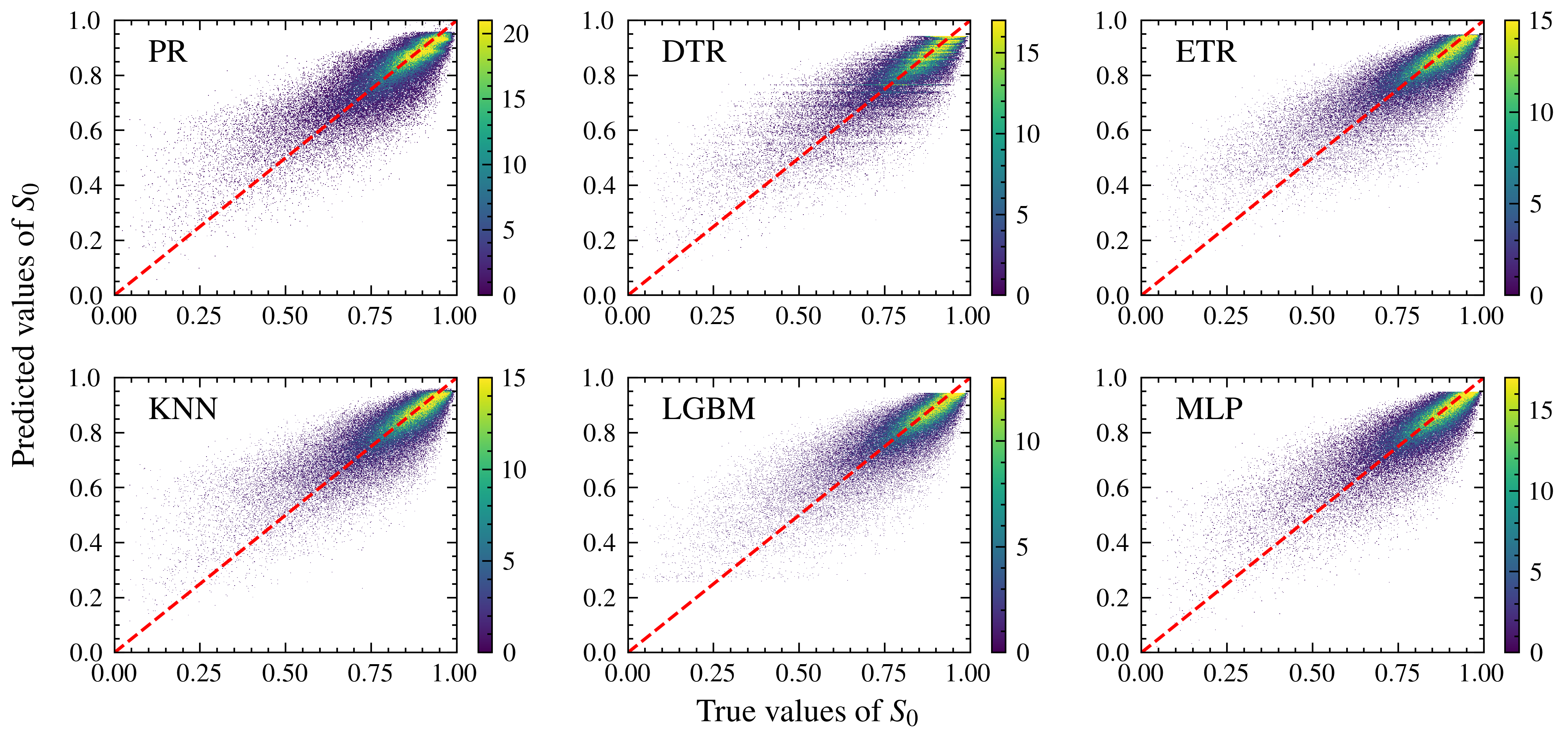}
\caption{ML models predictions versus true values of $S_0$ for 100 K testing data of minimum-bias Au+Au events at $\sqrt{s_{\rm NN}}=200$ GeV.}\label{fig3}
\end{figure}

\begin{figure}
\centering
\includegraphics[width=\textwidth]{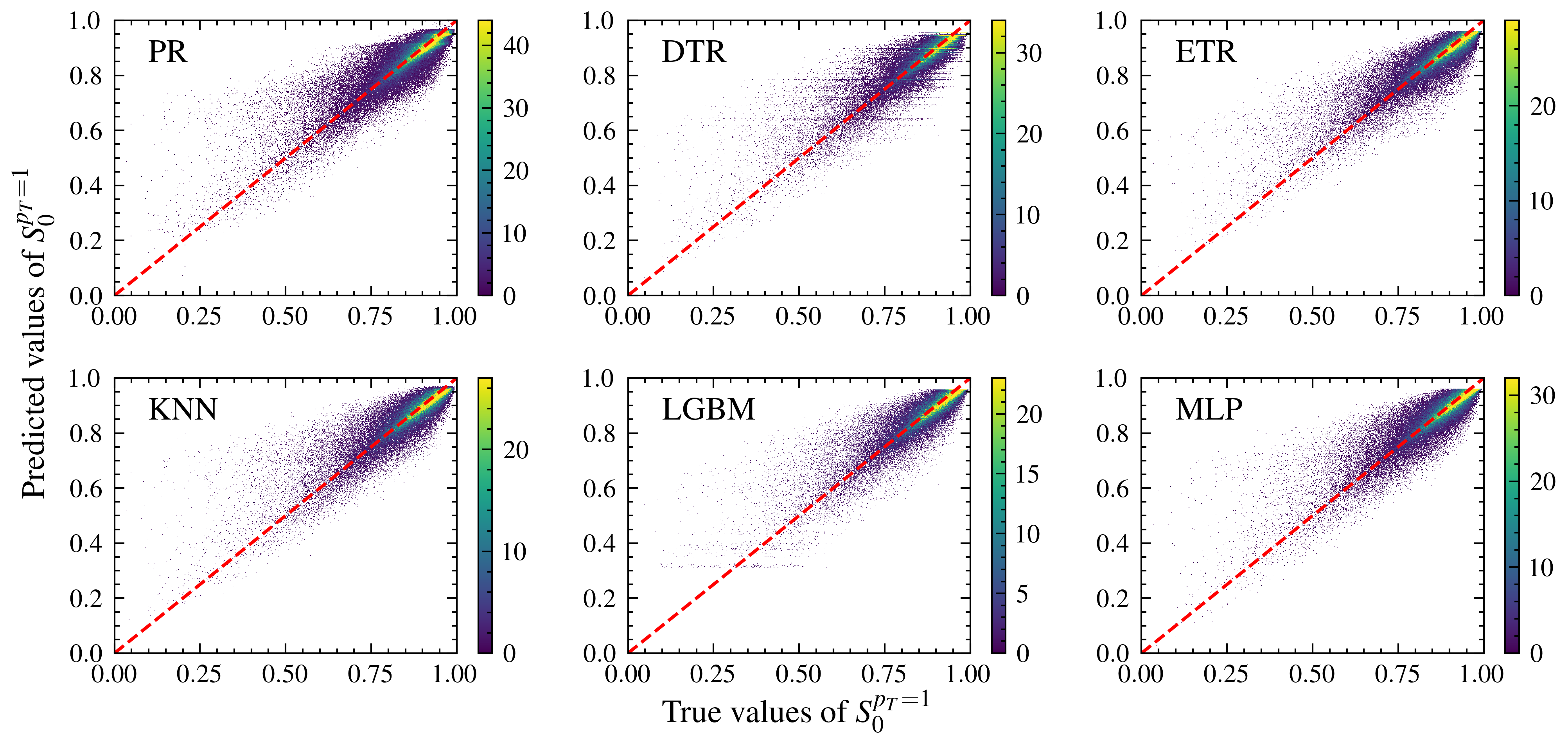}
\caption{ML models predictions versus true values of $S_0^{p_{\rm T}=1}$ for 100 K testing data of minimum-bias Au+Au events at $\sqrt{s_{\rm NN}}=200$ GeV.}\label{fig4}
\end{figure}

\begin{table}
\caption{Performance of the ML models in predicting $S_0$ values of minimum-bias Au+Au events at $\sqrt{s_{\rm NN}}=200$ GeV for training with 1 M events and testing with 100 K events. The statistical uncertainties were calculated by the bootstrap resampling method \cite{bootstrap}.}
\centering
\begin{tabular}{l c c c c}
\hline
\textrm{Model}& \textrm{MAE} & \textrm{RMSE} & $R^2$ & \textrm{Average Time (sec)}\\
\hline
PR & 0.04984 $\pm$ 0.00001 & 0.07118 $\pm$ 0.00001 & 0.69996 $\pm$ 0.00006 & 21.10 \\

DTR & 0.04960 $\pm$ 0.00003 & 0.07088 $\pm$ 0.00006 & 0.70250 $\pm$ 0.00047 & 3.27 \\

ETR & 0.05008 $\pm$ 0.00008 & 0.07094 $\pm$ 0.00008 & 0.70198 $\pm$ 0.00070 & 29.66 \\

KNN & 0.05075 $\pm$ 0.00004 & 0.07312 $\pm$ 0.00006 & 0.68340 $\pm$ 0.00048 & 3.84 \\

LGBM & 0.04929 $\pm$ 0.00001 & 0.07028 $\pm$ 0.00001 & 0.70753 $\pm$ 0.00011 & 7.83 \\

MLP & 0.04940 $\pm$ 0.00009 & 0.07040 $\pm$ 0.00014 & 0.70650 $\pm$ 0.00116 & 323.50 \\
\hline
\end{tabular}
\label{tab2}
\end{table}

\begin{table}
\caption{Performance of the ML models in predicting $S_0^{p_{\rm T}=1}$ values of minimum-bias Au+Au events at $\sqrt{s_{\rm NN}}=200$ GeV for training with 1 M events and testing with 100 K events.}
\centering
\begin{tabular}{l c c c c}
\hline
\textrm{Model}& \textrm{MAE} & \textrm{RMSE} & $R^2$ & \textrm{Average Time (sec)}\\
\hline
PR & 0.03363 $\pm$ 0.00001 & 0.05179 $\pm$ 0.00001 & 0.78013 $\pm$ 0.00006 & 19.94\\

DTR & 0.03339 $\pm$ 0.00002 & 0.05126 $\pm$ 0.00005 & 0.78466 $\pm$ 0.00040 & 3.07 \\

ETR & 0.03379 $\pm$ 0.00010 & 0.05151 $\pm$ 0.00010 & 0.78254 $\pm$ 0.00089 & 29.86 \\

KNN & 0.03424 $\pm$ 0.00003 & 0.05323 $\pm$ 0.00005 & 0.76777 $\pm$ 0.00041 & 3.63 \\

LGBM & 0.03330 $\pm$ 0.00001 & 0.05107 $\pm$ 0.00001 & 0.78622 $\pm$ 0.00012 & 1.58\\

MLP & 0.03353 $\pm$ 0.00014 & 0.05144 $\pm$ 0.00027 &  0.78315 $\pm$ 0.00225 &  142.79 \\
\hline
\end{tabular}
\label{tab3}
\end{table}

\section{\label{Results}Results and Discussion}

\begin{figure}[tb]
\centering
\includegraphics[width=\textwidth]{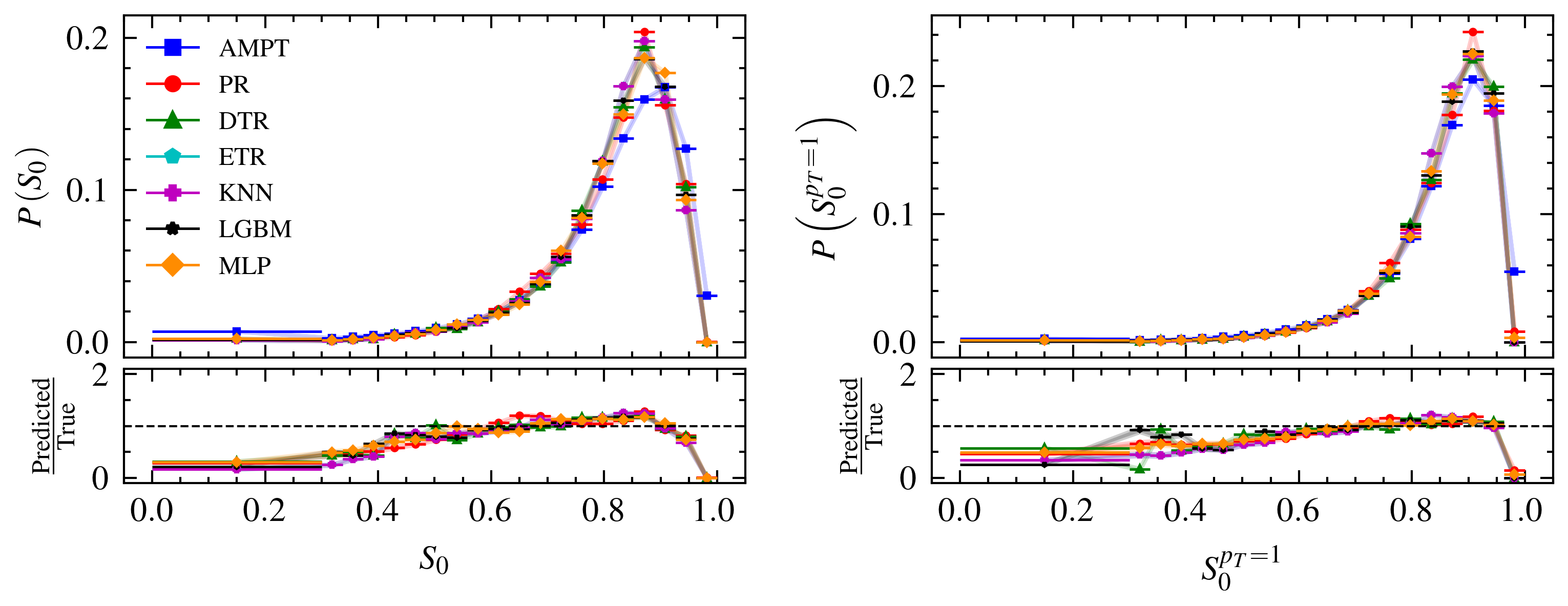}
\caption{Comparison between the ML-predicted and true distributions of conventional spherocity (left) and unweighted spherocity (right) for minimum-bias Au+Au events at $\sqrt{s_{\rm NN}}=200$ GeV. The ratio of the model predicted values to the true values of spherocity are shown in the lower panels.} \label{fig5}
\end{figure}

All machine learning models, after hyperparameter optimization, were trained on one million minimum-bias \texttt{AMPT}-generated Au+Au events at $\sqrt{s_{\rm NN}}=200$ GeV, and their performance was assessed using an independent test set of 100 K \texttt{AMPT} events at the same collision energy. Correlation plots comparing the predicted values of $S_0$ and $S_0^{p_{\rm T}=1}$ for all ML models with their true values are presented in Figures \ref{fig3} and \ref{fig4}.
These plots indicate that the predicted and true values generally follow the expected trend represented by the red dashed line. Nevertheless, noticeable deviations appear in the jetty region, particularly at lower values of $S_{0}$ and $S_{0}^{p_{\rm T}=1}$. To better understand the model performance, we compare the ML-predicted spherocity (both traditional and unweighted) values with the true values from the AMPT model, as shown in Fig.~\ref{fig5}. The lower panels display the corresponding ratios. The figure shows a systematic reduction in predictive performance for all ML models toward low $S_0$, reflecting the increased difficulty in describing jetty topologies. Furthermore, the sharp drop in performance at very small $S_0$ may be attributed to the low production cross-section of highly jetty events, which results in very limited training statistics for these rare topologies. The performance of the machine learning models is quantitatively presented in Tables~\ref{tab2} and \ref{tab3}. The analysis of various machine learning models reveals that unweighted spherocity achieves systematically better predictions than the traditional definition across all metrics. It achieves lower MAE and RMSE and higher $R^{2}$, reflecting its ability to characterize the global event shape more effectively for ML learning. This improvement arises because the unweighted definition eliminates the bias associated with high-$p_{\rm T}$ particles that is inherent in the traditional formulation. Among the models tested, LGBM provides the best overall performance, combining the lowest errors and highest $R^{2}$ with very short execution time, while DTR also achieves competitive accuracy with minimal runtime. In contrast, MLP attains accuracy comparable to LGBM but at substantially higher computational cost, and KNN shows weaker predictive power. Importantly, the present approach demonstrates a marked improvement over the earlier study \cite{Mallick_sph_ML}. All models were trained on a 4-core, 8-thread Intel(R) Core(TM) i5-10300H CPU with 16 GB DDR4 RAM.

\begin{figure}[tb]
\centering
\includegraphics[width=\textwidth]{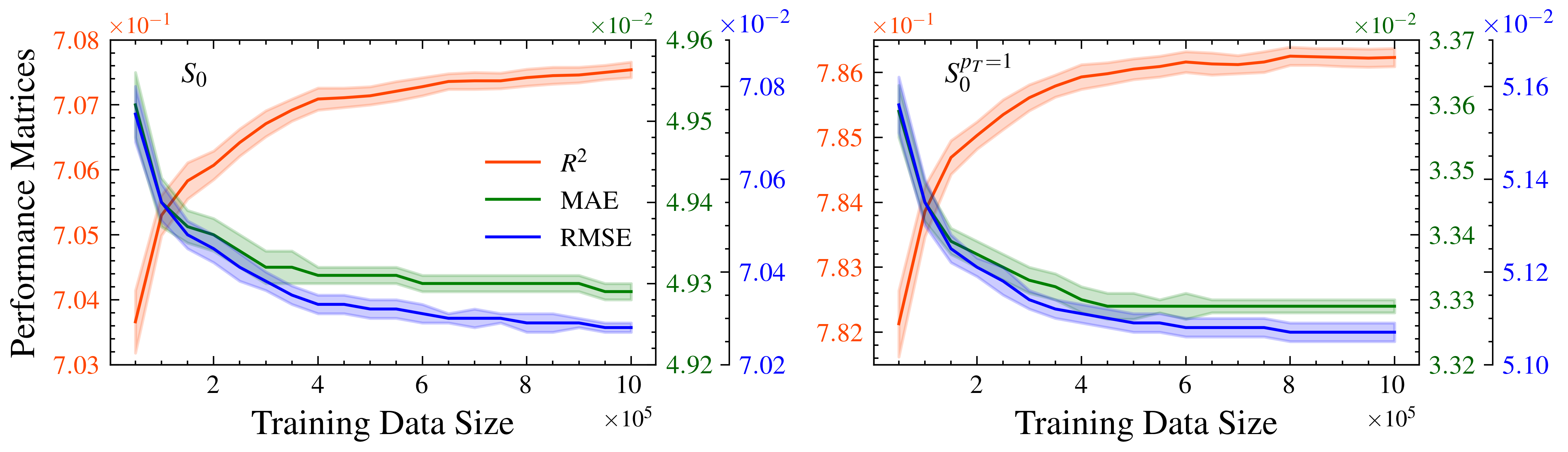}
\caption{Size dependence of the performance of the LGBM model with 
training data size in predicting conventional spherocity (left) and unweighted spherocity (right) for 100 K minimum-bias Au+Au test events at $\sqrt{s_{\rm NN}}=200$ GeV.}\label{fig6}
\end{figure}

To investigate the dependence of model performance on the size of the training dataset, we evaluated the LGBM model using different training set sizes, while the test set was fixed at 100 K events. Results, shown in Fig.~\ref{fig6}, indicate a marginal improvement with larger training sizes, with performance stabilizing beyond 100~K training samples. This indicates that 100~K training data are already sufficient for near-optimal model performance, with additional data primarily improving convergence and reducing statistical fluctuations. Since our ML models were trained with one million events, this dataset is considered sufficiently large and close to optimal.

\subsection{Effect of Reaction plane angle}

The elliptic flow coefficient $v_{2}$, used as an input to the machine-learning models, depends explicitly on the reaction plane angle $\psi$. While $\psi$ is available in \texttt{AMPT} simulations, it is not directly measurable in experiment; in its absence, $\langle \cos(2\phi) \rangle$ can be considered as a proxy for $v_{2}$. To investigate the effect of reaction plane angle information on model performance, we consider three scenarios. Scenario~I (ideal): the LGBM model is trained and tested with $\psi$ known (Tables~\ref{tab2}--\ref{tab3}). Scenario~II: the model is trained on data containing the true $v_{2}$ ($\psi$ known) but tested on data where $\psi$ is absent (Table~\ref{tab4}). Scenario~III (realistic): the model is both trained and tested without $\psi$ (Table~\ref{tab5}). The results show that Scenario~I achieves the highest predictive accuracy, Scenario~III yields the next best performance, and Scenario~II exhibits a noticeable degradation. These observations indicate that consistency between training and testing feature sets with respect to $\psi$ reduces performance loss when applying models to experimental ($\psi$-absent) data.\\

\begin{table}
\caption{Performance of the LGBM model in predicting $S_0$ and $S_0^{p_{\rm T}=1}$ values of minimum-bias Au+Au events at $\sqrt{s_{\rm NN}}=200$ GeV. Here the model is trained on data containing the true $v_2$ ($\psi$ known) but tested on data where $\psi$ is absent.}
\centering
\begin{tabular}{l c c }
\hline
\textrm{Observable} & $S_0$ & $S_0^{p_{\rm T}=1}$ \\ 
 \hline
MAE & 0.06140 $\pm$ 0.00003 & 0.04707 $\pm$ 0.00003\\
RMSE & 0.08038 $\pm$ 0.00003 & 0.06178 $\pm$ 0.00003\\
$R^2$ & 0.61740 $\pm$ 0.00026 & 0.68721 $\pm$ 0.00032\\
\hline
\end{tabular}
\label{tab4}
\end{table}

\begin{table}
\caption{Performance of the LGBM model in predicting $S_0$ and $S_0^{p_{\rm T}=1}$ values of minimum-bias Au+Au events at $\sqrt{s_{\rm NN}}=200$ GeV. Here the model is both trained and tested without $\psi$.}
\centering
\begin{tabular}{l c c }
\hline
\textrm{Observable} & $S_0$ & $S_0^{p_{\rm T}=1}$ \\ 
 \hline
MAE & 0.05302 $\pm$ 0.00001 & 0.03757 $\pm$ 0.00001\\
RMSE & 0.07360 $\pm$ 0.00001 & 0.05442 $\pm$ 0.00002\\
$R^2$ & 0.67924 $\pm$ 0.00013 & 0.75730 $\pm$ 0.00017\\
\hline
\end{tabular}
\label{tab5}
\end{table}

\begin{table}
\caption{Comparison of the performance of the LGBM model in predicting $S_0$ values of minimum-bias Au+Au events at $\sqrt{s_{\rm NN}}=200$ GeV trained with 1 M \texttt{AMPT} data and tested with 50 K data of different event generators.}
\centering
\begin{tabular}{l c c c}
\hline
\textrm{Event Generator}& \textrm{MAE}& \textrm{RMSE } & \textrm{$R^2$ } \\
 \hline
\texttt{AMPT} & 0.05319 $\pm$ 0.00002 & 0.07402 $\pm$ 0.00003 & 0.67433 $\pm$ 0.00023\\
\texttt{UrQMD} & 0.05554 $\pm$ 0.00005 & 0.07558 $\pm$ 0.00007 & 0.63144 $\pm$ 0.00070 \\
\texttt{PYTHIA8 Angantyr} & 0.05783 $\pm$ 0.00008 & 0.08201 $\pm$ 0.00009 & 0.71263 $\pm$ 0.00065\\
\texttt{EPOS4} & 0.06562 $\pm$ 0.00011 & 0.10030 $\pm$ 0.00022 & 0.60938 $\pm$ 0.00175 \\
\hline
\end{tabular}
\label{tab6}
\end{table}

\begin{table}
\caption{Comparison of the performance of the LGBM model in predicting $S_0^{p_{\rm T}=1}$ values of minimum-bias Au+Au events at $\sqrt{s_{\rm NN}}=200$ GeV trained with 1 M \texttt{AMPT} data and tested with 50 K data of different event generators.}
\centering
\begin{tabular}{l c c c}
\hline
\textrm{Event Generator} &\textrm{MAE} &\textrm{RMSE} & \textrm{$R^2$} \\
\hline
\texttt{AMPT} & 0.03791 $\pm$ 0.00002 & 0.05484 $\pm$ 0.00003 & 0.75318 $\pm$ 0.00025 \\
\texttt{UrQMD} & 0.04017 $\pm$ 0.00004 & 0.05683 $\pm$ 0.00005 & 0.70718 $\pm$ 0.00056 \\
\texttt{PYTHIA8 Angantyr} & 0.04167 $\pm$ 0.00005 & 0.06479 $\pm$ 0.00009 & 0.77251 $\pm$  0.00062 \\
\texttt{EPOS4} & 0.04722 $\pm$ 0.00007 & 0.07819 $\pm$ 0.00016 & 0.68776 $\pm$ 0.00131 \\
\hline
\end{tabular}
\label{tab7}
\end{table}

\begin{table}[h]
\caption{omparison of the performance of the LGBM model in predicting $S_0$ values of minimum-bias Au+Au events at $\sqrt{s_{\rm NN}}=200$ GeV trained with 600 K \texttt{PYTHIA8 Angantyr} data and tested with 50 K data of different event generators.}
\centering
\begin{tabular}{l c c c}
\hline
\textrm{Event Generator}& \textrm{MAE}& \textrm{RMSE } & \textrm{$R^2$ } \\
 \hline
\texttt{AMPT} & 0.06118 $\pm$ 0.00010 & 0.08700 $\pm$ 0.00024 & 0.55006 $\pm$ 0.00245 \\
\texttt{UrQMD} & 0.04499 $\pm$ 0.00004 & 0.06765 $\pm$ 0.00008 & 0.70477 $\pm$ 0.00070 \\
\texttt{PYTHIA8 Angantyr} & 0.04620 $\pm$ 0.00002 & 0.06966 $\pm$ 0.00003 & 0.79264 $\pm$ 0.00017 \\
\texttt{EPOS4}& 0.05733 $\pm$ 0.00019 & 0.09791 $\pm$ 0.00044 & 0.62778 $\pm$ 0.00331 \\
\hline
\end{tabular}
\label{tab8}
\end{table}

\begin{table}[h]
\caption{Comparison of the performance of the LGBM model in predicting $S_0^{p_{\rm T}=1}$ values of minimum-bias Au+Au events at $\sqrt{s_{\rm NN}}=200$ GeV trained with 600 K \texttt{PYTHIA8 Angantyr} data and tested with 50 K data of different event generators.}
\centering
\begin{tabular}{l c c c}
\hline
\textrm{Event Generator} &\textrm{MAE} &\textrm{RMSE} & \textrm{$R^2$} \\
\hline
\texttt{AMPT} & 0.04353 $\pm$ 0.00011 & 0.06615 $\pm$ 0.00023 & 0.64088 $\pm$ 0.00250 \\
\texttt{UrQMD} & 0.03323 $\pm$ 0.00003 & 0.05193 $\pm$ 0.00007 & 0.75549 $\pm$ 0.00063 \\
\texttt{PYTHIA8 Angantyr} & 0.03274 $\pm$ 0.00001 & 0.05353 $\pm$ 0.00002 & 0.84471 $\pm$ 0.00013 \\
\texttt{EPOS4} & 0.04013 $\pm$ 0.00011 & 0.07637 $\pm$ 0.00028 & 0.70217 $\pm$ 0.00215 \\
\hline
\end{tabular}
\label{tab9}
\end{table}

\subsection{Event Generator dependency}

One limitation of machine learning algorithms is their dependence on the event generator used for training. A model trained on simulated data from a specific generator may not yield accurate predictions when applied to data from a different generator, owing to differences in the underlying physics mechanisms. To investigate this dependence, we examine whether our ML model, trained on \texttt{AMPT}-simulated data, can reliably predict spherocity when tested on data from other event generators. For this purpose, we employ \texttt{UrQMD}, \texttt{PYTHIA8 Angantyr}, and \texttt{EPOS4}. Since these generators do not provide the reaction plane angle, we use $\langle \cos{(2\phi)}\rangle$ instead of elliptic flow both in the training and test datasets. The LGBM model is trained using $1\,\mathrm{M}$ \texttt{AMPT}-simulated Au+Au collision events at $\sqrt{s_{_{\mathrm{NN}}}} = 200~\mathrm{GeV}$ and tested on $50\,\mathrm{K}$ events from each of the other generators at the same energy, predicting both conventional and unweighted spherocity. We have further trained the LGBM model on 600 K \texttt{PYTHIA8 Angantyr} events and tested it on 50~K events from each of the other studied event generators for the prediction of both the conventional and unweighted spherocity. The results, shown in Tables~\ref{tab6}--\ref{tab9}, display consistently good performance across all test sets. This indicates that the correlation between the global observables $dN_{\rm ch}/d\eta$, $\langle m_T \rangle$, and $v_2$ with $S_0$ is largely generator independent. Consequently, the trained ML models exhibit reduced event-generator dependence, supporting their applicability to experimental data.

\subsection{SHAP-based Interpretation of Model Predictions}

\begin{figure}[h!]
\centering
\includegraphics[width=0.8\textwidth]{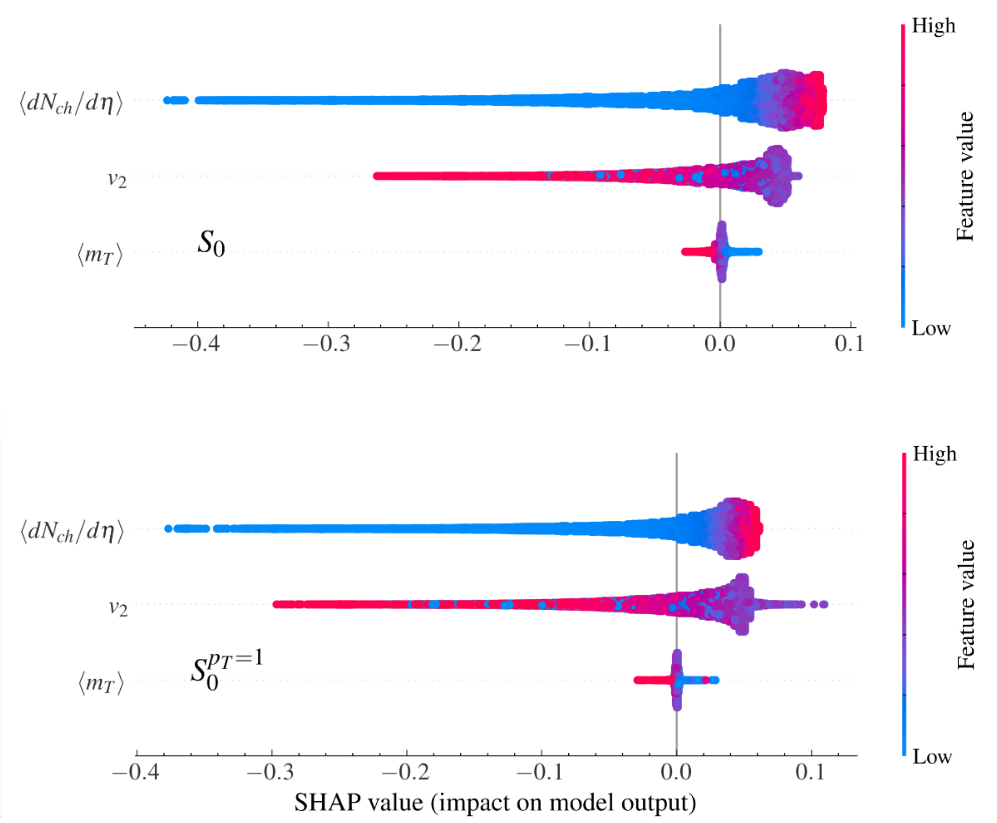}
\caption{SHAP summary plots from the LGBM model for the prediction of $S_0$ (top) and $S_0^{p_{\rm T}=1}$ (bottom).}\label{fig7}
\end{figure}

Machine learning (ML) models are often regarded as ‘black boxes’ because the mechanisms underlying their predictions are not readily accessible. To assess the influence of individual input features on the predictions, we employ the SHapley Additive exPlanations (SHAP) method \cite{SHAP}, which, based on cooperative game theory, expresses the model output as the sum of Shapley values for each feature. This allows quantification of feature importance and identification of the most influential inputs.
Figure~\ref{fig7} displays the SHAP summary plots for the LGBM regressor in estimating $S_{0}$ and $S_0^{p_{\rm T}=1}$. Features are ranked vertically by importance, with SHAP values on the horizontal axis indicating the magnitude and direction of their contribution. Point colors denote feature values. Positive SHAP values indicate that a feature increases the predicted output, whereas negative values indicate a decreasing effect. The results show that charged-particle multiplicity $\langle dN_{\mathrm{ch}}/d\eta \rangle$ has the largest positive impact, followed by elliptic flow $v_{2}$, while transverse mass $\langle m_{\rm T} \rangle$ contributes least. The former shows a strong positive correlation with the output, whereas $v_{2}$ and $\langle m_{\rm T} \rangle$ exhibit negative correlations.

\section{\label{Summary}Summary}

We have applied several machine-learning (ML) algorithms to identify isotropic and jet-like events in Au+Au collisions at $\sqrt{s_{\rm NN}} = 200$ GeV, simulated with the \texttt{AMPT} model. The ML models were trained on event-level raw observables related to event topology, including charged-particle multiplicity at midrapidity $\langle dN_{\mathrm{ch}}/d\eta \rangle$, average transverse mass $\langle m_{\rm T} \rangle$, and elliptic flow $\langle v_{2} \rangle$, to predict both conventional spherocity $S_{0}$ and unweighted transverse spherocity $S_0^{p_{\rm T}=1}$. Hyperparameter optimization was performed for each model, with the LGBM algorithm yielding the best performance. Predictions of unweighted spherocity were generally more accurate than those of conventional spherocity, although performance degraded in jetty events for both cases.

To assess robustness with realistic inputs, $v_{2}$ was replaced by $\langle \cos(2\phi) \rangle$ to remove explicit reaction plane information, resulting in only a slight reduction in performance, indicating sensitivity to the reaction plane is weak. Generalizability tests using data from \texttt{UrQMD}, \texttt{PYTHIA8 Angantyr}, and \texttt{EPOS4} showed only minor performance loss, suggesting reduced event-generator dependence. 
These results point to a potential application of the trained ML models to experimental data for spherocity estimation.

\data{The simulation data that support the findings of this study are available upon reasonable request from the authors.}

\providecommand{\noopsort}[1]{}\providecommand{\singleletter}[1]{#1}%

\end{document}